\documentclass[%
reprint,
showpacs,
preprintnumbers,
showkeys,
amsmath,amssymb,
prd,
nofootinbib,
]{revtex4-1}

\RequirePackage[T1]{fontenc}

\usepackage{amssymb,amsmath,latexsym}

\RequirePackage{graphicx}
\RequirePackage{mathptmx}      
\RequirePackage{flushend}
\RequirePackage[colorlinks,citecolor=blue,urlcolor=blue,linkcolor=blue]{hyperref}

\usepackage{sidecap}
\usepackage{ifthen}
\usepackage{keycommand}
\usepackage[toc,page]{appendix}
\usepackage[labelfont={small,bf},textfont=small]{caption}
\usepackage[labelfont={small,bf},textfont=small]{subcaption}

\usepackage{placeins} 

\makeatletter \let\cl@chapter\relax \makeatother 

\usepackage{cleveref}  
\newcommand{\refp}[1]{(\ref{#1})}

\crefname{equation}{eq.}{eqs.}
\crefname{section}{sect.}{sects.}
\crefname{chapter}{chapter}{chapters}
\crefname{table}{table}{tables}
\crefname{figure}{fig.}{figs.}
\crefname{appsec}{appendix}{appendices}
\crefname{appchap}{appendix}{appendices}

\DeclareMathOperator{\e}{e}
\makeatletter
\def\blfootnote{\xdef\@thefnmark{}\@footnotetext}
\makeatother

\DeclareCaptionJustification{justified}{\justifying}
\newcommand{\TITLE}{Hamilton equations and inertial mass increase}

\begin{document}
\title{\TITLE}

\author{M. V. Lokaj\'{\i}\v{c}ek}
\email{lokaj@fzu.cz}
\author{Ji\v{r}\'{\i} Proch\'{a}zka}
\email{jiri.prochazka@fzu.cz}
\affiliation{The Czech Academy of Sciences, Institute of Physics}


\begin{abstract}
It has been shown in the past century that the particle inertia against velocity change has increased at higher velocity values. This increase has been predicted in principle in the framework of special theory of relativity. 
However, any comparison of the corresponding prediction with experimental data obtained already in the first half of the past century has not been provided until now.
It will be shown in the presented paper that quite arbitrary inertia mass increase with velocity may be described also in the framework of the classical physics on the basis of Hamilton's equations if the force law of Newton will be generalized; i.e., if time change of particle momentum (not directly acceleration) will be determined by corresponding force. More general velocity-dependent formulas (containing some free parameters) for kinetic energy, momentum and force will be then derived.
It will be further shown that this generalized Hamiltonian mechanics describing general mass increase with velocity may be reduced to known result of classical physics of Newton if mass value is taken to be constant, and also that the result of special theory of relativity may be derived if the given increasing function is taken as predicted in this theory.
\end{abstract}

\pacs{45.20.-d,03.30.+p,29.20.-c}
\keywords{Hamilton's equations, inertia mass increase, accelerated particles}

\maketitle

\section{Introduction} 

The existence of particle inertia increase with velocity was predicted by special theory of relativity (STR) in the past century (see \cite{Einstein1905_relativity,Einstein1905_inertia}). However, any comparison with corresponding  experimental data has not been presented until now, even if they have existed since the first half of the past century.
It will be shown in the following that similar phenomenon may be derived also in the framework of generalized classical physics (GCP) when Newton's force law will be correspondingly generalized; i.e., the effect of a force on a particle will be defined as $\vec{F} = \text{d}\vec{p}/\text{d}t = \text{d}(m(v)\,\vec{v})/\text{d}t$.
The basic equations of Hamilton may remain without any change in such a case. It is necessary only to admit in the definition of Hamiltonian that the corresponding mass values of individual particles are not constant; they are  to be defined as dependent on corresponding velocity values (possibly in dependence on some free parameters).

The generalized classical system involving very \emph{arbitrary} increase of inertial mass with velocity will be introduced in 
\cref{sec:general_system}. 
The GCP solutions may be rather diverse; they must be always limited only at very low velocity (kinetic energy) values to be in agreement with all results of classical mechanics. 
It will be then shown in \cref{sec:alternatives} that the solutions may be reduced not only to the results known from classical physics of Newton, but also to those predicted by special theory of relativity.
In \cref{sec:comparison} some different possible alternatives will be presented and compared to the behavior obtained in the framework of special theory of relativity.

\section{\label{sec:general_system}General classical system}
 
Let us start with the motion of a matter particle in a force field. The effect of force (represented by vector $\,\vec{F}\;$) on a particle having position $\vec{q}$, velocity $\,\vec{v}\;$ and momentum $\vec{p}$ may be characterized in the three-dimensional space by following conditions ($v=|\vec{v}|$, similar notation will be used for any other vector and its magnitude):
\begin{subequations}
\label{eq:basic_conditions}
\begin{align}   
\vec{p}\,=&\,m(v)\,\vec{v} \, , \label{eq:p_v} \\
\vec{F}\, \text{d}t \, =& \,\text{d}\vec{p}\, ,     \label{eq:F_dpdt} \\
\text{d}\vec{q} \,=& \,\vec{v}\,\text{d}t \, ,      \\
\vec{F}.\text{d}\vec{q} \,=& \, \text{d}E_{\text{kin}}(v) \,,  \label{eq:dEkindv}     
\end{align}
\end{subequations}
where $\;E_{\text{kin}}(v)\;$ represents corresponding kinetic energy value  and $\;m(v)\;$ represents the inertia of a given particle (i.e., the ability of a given force to change particle velocity at a given velocity value). The space-coordinates effect of force represents the increment of work; the time effect is given by the change of momentum (not directly by acceleration).

The formula \refp{eq:F_dpdt} reduces to standard Newton law when $\;m(v)\,=\,m(v\!=\!0)\,$. In generalized case it follows from the preceding equations  

\begin{equation}     
\frac{\text{d}E_{\text{kin}}(v)}{\text{d}v} \;=\;\frac{\text{d}\vec{p}}{\text{d}v}\,.\,\frac{\text{d}v}{\text{d}t}\,\frac{\text{d}\vec{q}}{\text{d}v}\;=\;\frac{\text{d}\vec{p}(v)}{\text{d}v}\,.\, \vec{v}   \;,       \label{dEv}
\end{equation}
\begin{equation}  
    \frac{\text{d}E_{\text{kin}}(v)}{\text{d}v} \;=\;  \vec{v}.\left( m(v)\frac{\text{d}\vec{v}}{\text{d}v} \,+\,\vec{v}\,\frac{\text{d}m(v)}{\text{d}v}\right) \\
                                       \,=\, v\;m(v)  \,+\, v^2\,\frac{\text{d}m(v)}{\text{d}v}       \; ;   \label{eq:dEdv}
\end{equation}
it holds also 
\begin{equation} 
   \frac{\text{d}E_{\text{kin}}}{\text{d}t}= \vec{F}.\vec{v} \;.                 \label{eq:dE_kin_dt}
\end{equation}
Integrating \cref{eq:dEdv} from $\,0$ to $v\,$ following general relation for kinetic energy of accelerated particle to velocity $v$
\begin{equation}
   E_{\text{kin}}(v) \,=\, \left( v^2 m(v) - \int_0^v \text{d}v' \, v' m(v') \right)     
\label{eq:E_kin_v}
\end{equation}
may be derived. 
On the basis of \cref{eq:p_v,eq:F_dpdt} one may write
\begin{equation}  
  \vec{F} =  \left(\frac{\text{d}\vec{v}}{\text{d}t} m(v)\,+\vec{v}\frac{\text{d}m(v)}{\text{d}v} \frac{\text{d}v}{\text{d}t}\right)\, .
 \label{eq:F_v}
\end{equation}

Instead of $\,\vec{v}\,$ we may define dimensionless vector and its magnitude (as it has been done also in the context of STR)
\begin{equation}
\vec{\beta} = \frac{1}{c} \vec{v} \, , \;\;\;\;\; \beta = \frac{v}{c}
\label{eq:beta}
\end{equation}
where constant $\,c\,$ represents the speed of light in vacuum (suitably chosen velocity unit to simplify the solution of given problem). Variable $\vec{\beta}$ (or its magnitude $\beta$) will be called also velocity in the following for the sake of simplicity (there is always one-to-one correspondence between $\vec{v}$ and $\vec{\beta}$).

To describe inertia increase we shall further define for any moving particle the function
\begin{equation}
     m(\beta)\;=\; m_0\,f(\beta)      
\label{eq:m_beta}
\end{equation}
where $\,f(\beta)\,$ is a dimensionless monotonically rising function of velocity and parameter $m_0\;=\;m(\beta\!=\!0)$ (rest mass) may be different for each particle.

The motion is to correspond to the results of classical mechanics in the region of very low velocity values; it is necessary to hold 
\begin{subequations}
\label{eq:f_conditions}
\begin{align}
 f(\beta\!=\!0) \,&=\, 1, \\ 
 \frac{\text{d}f}{\text{d}\beta}(\beta\!=\!0) \,&=\, 0 , \\
\frac{\text{d}f}{\text{d}\beta}(\beta) \; &\ge\, 0\,\,.
\end{align}
\end{subequations}

The monotonically rising function $\;f(\beta)\;$ is then to fulfill also the conditions derived from basic classical equations \refp{eq:basic_conditions}; it should hold (see \cref{eq:beta,eq:m_beta} and \cref{eq:E_kin_v,eq:p_v,eq:F_dpdt})

\begin{align}
E_{\text{kin}}(\beta) \,=&\, m_0c^2 \left( \beta^2 f(\beta) - \int_0^\beta \text{d}\beta' \, \beta' f(\beta') \right) \, ,  \label{eq:E_kin_beta} \\
\vec{p} \,=&\, m_0  f(\beta) \vec{\beta}\, , \label{eq:p_beta} \\
\vec{F} \,=&\, m_0\,c\left(\frac{\text{d}\vec{\beta}}{\text{d}t} f(\beta)\,+\vec{\beta}\frac{\text{d}f(\beta)}{\text{d}\beta} \frac{\text{d}\beta}{\text{d}t}\right)\, .  \label{eq:F_beta}
\end{align}

The general formulas \refp{eq:E_kin_beta}, \refp{eq:p_beta} and \refp{eq:F_beta} have been derived for \emph{arbitrary} function $f(\beta)$ (limited only by the conditions \refp{eq:f_conditions}) and, therefore, might allow to introduce quite different dependences of inertia increase specified by function $f(\beta)$. Parameter $m_0$ may be determined in principle from the data of classical physics at very low velocities. The actual shape of function $f(\beta)$ (which may contain some additional free parameters, too) should be, however, derived from data at higher velocity values.

The given approach will hold for any physically acceptable definition of the potentials between individual particles. In extreme case one particle may be substituted by some facilities providing corresponding (eventually constant) force in the whole considered space. Effect of this force on particle motion, at different velocity values, should be measured. It concerns mainly electromagnetic force acting on charged particle. The function $f(\beta)$ (containing some free parameters) can be then derived from experimental data with the help of \cref{eq:F_beta}.

The time evolution of a given particle system (of any individual coordinate) may be described with the help of standard Hamilton's equations ($q_i,\, p_i\;$ representing individual components of corresponding vectors; see e.g. \cite{Arnold1989}) 
\begin{eqnarray}
   \frac{\text{d} q_i}{\text{d}t} =    \frac{\partial H}{\partial p_i},    \;\;\;\;\;\;\;\;\;
      \frac{\text{d} p_i}{\text{d}t} =    -\frac{\partial H}{\partial q_i}        \label{eq:hamilton}
\end{eqnarray}
where the Hamiltonian $H$ representing the total energy (the sum of kinetic energy and corresponding potential energy for all particles) and the force acting on a corresponding particle equal ($p=|\vec{p}|$)
\begin{equation}  
   H = E_{\text{kin}}(p) +U(\vec{q}) , \;\;\;\;\;\;  \vec{F}\,=\,-\nabla U(\vec{q})\,.
\end{equation} 
This force may be derived from the function $U(\vec{q})$ (the sum of potentials acting from other particles). The expression for kinetic energy (as well as momentum) of individual particles is to be generalized in agreement with relations \refp{eq:E_kin_beta}, \refp{eq:p_beta} and \refp{eq:F_beta}.

The derived general formulas may describe very different physical characteristics (very different possibilities). In the following we introduce for demonstration purposes some different alternatives which are analytically solvable corresponding to physical systems with infinite as well as finite maximal velocity (the general formulas do not contain any a priory limit on the maximal velocity value $\beta_{\text{max}}$).

\section{\label{sec:alternatives}Alternative possibilities of inertia increase}

\subsection{\label{sec:transition_newton}Transition to Newton's theory}
First of all let us start with showing that the general relations \refp{eq:E_kin_beta} and \refp{eq:p_beta} may reproduce the results known from the classical physics.  If 

\begin{equation}    
         f(\beta) \;=\;  1,
\label{eq:f_v_newton}
 \end{equation}
i.e., $m(\beta)$ defined by \cref{eq:m_beta} does not depend on velocity, then the kinetic energy given by the general formulas may be written in the form

\begin{equation}  
   E_{\text{kin}}(\beta) \,=\, \frac{1}{2}m_0 c^2 \,\beta^2 = \, \frac{1}{2}m_0 \,v^2
\label{eq:E_kin_newton}
\end{equation}  
\begin{equation}
       \vec{p}   \,= \, m_0 \, c\vec{\beta} =\, m_0 \, \vec{v} , 
\end{equation}
which are well known relations of classical physics of Newton.

\subsection{\label{sec:infinite_velocity_limit}Infinite velocity limit}  
In the previous case the maximum particle velocity $\beta_{\text{max}}$ might tend to infinity. However, it may occur in many different ways; as an example, we may introduce simple exponential function with one free parameter ($\epsilon>0$)

\begin{equation}
   f(\beta) \,=\, \e^{\frac{1}{2}(\frac{\beta}{\epsilon})^2}  
\label{eq:f_v_exponential}
  \end{equation}
which satisfies evidently general conditions \refp{eq:f_conditions} and which may describe the mass increase defined by \cref{eq:m_beta}. On the basis of \cref{eq:E_kin_beta} one then obtains for kinetic energy
\begin{equation}
      E_{\text{kin}}(\beta) \,=\, m_0\, (\epsilon c)^2
   \left[\e^{\frac{1}{2}(\frac{\beta}{\epsilon})^2}\left(\left(\frac{\beta}{\epsilon}\right)^2-1\right)+1\right]   \, ,
\label{eq:E_kin_exponential}
\end{equation}
momentum being given by \cref{eq:p_beta}.

With the help of \cref{eq:E_kin_exponential,eq:p_beta} and Taylor expansions at zero velocity one may write
\begin{equation}
\begin{split}
E_{\text{kin}}(\beta) \,\approx&\frac{1}{2}m_0\,c^2 \beta^2 \,  \\
&\left[ 1  + \frac{3}{4}\left(\frac{\beta}{\epsilon}\right)^{2} + \frac{5}{24}\left(\frac{\beta}{\epsilon}\right)^{4}  + \mathcal{O}\left(\left(\frac{\beta}{\epsilon}\right)^{6}\right)  \right]\, ,
\end{split}
\label{eq:E_kin_exponential_taylor}
\end{equation}
\begin{equation} 
\vec{p} \,\approx\, m_0\, c\vec{\beta}\left[1+ \frac{1}{2}\left(\frac{\beta}{\epsilon}\right)^{2} + \frac{1}{8}\left(\frac{\beta}{\epsilon}\right)^{4} + \mathcal{O}\left(\left(\frac{\beta}{\epsilon}\right)^{6}\right) \right] \, .
\label{eq:p_exponential_taylor}
\end{equation}

The agreement with classical results, see 
\cref{sec:transition_newton}, 
will be conserved in the region of velocity values in which holds $\; v \ll \epsilon c$ (i.e., $\beta \ll \epsilon$). Very different behavior exists, however, in the region of great velocity values for different values of parameter $\,\epsilon\,$. In all these cases the velocity values increase to infinity with rising kinetic energy value (or vice versa). Several different cases will be shown in 
\cref{sec:comparison}. 

\subsection{\label{sec:finite_velocity_limit}Finite velocity limit}  
It is possible, of course, to define the increasing function $\,f(\beta)\,$ in different ways, too. Instead of exponential form it may be chosen, e.g.,
\begin{equation} 
f(\beta) \,= \,\frac{1}{\sqrt{1\,-\,(\frac{\beta}{\epsilon})^2}}\,      
\label{eq:f_v_str_epsilon}
 \end{equation} 
where $\epsilon$ is again a free parameter greater then zero. The kinetic energy will go to infinity at maximal velocity value 
\begin{equation}
\beta_{\text{max}}\,=\, \epsilon
\label{eq:v_max_str_epsilon}
\end{equation}
(it must hold $\;\beta<\beta_{\text{max}}\,$). 
It will hold then further (see \cref{eq:E_kin_beta})
\begin{equation}
   E_{\text{kin}}(\beta)\,= m_0 c^2 \epsilon^2 \left( f(\beta) - 1 \right)   \,.
\label{eq:E_kin_str_epsilon}
\end{equation}

If 
$\epsilon\!=\!1$, then \cref{eq:f_v_str_epsilon,eq:v_max_str_epsilon,eq:E_kin_str_epsilon,eq:p_beta} correspond fully to the result derived in the framework of STR in the previous century.
 
With the help of \cref{eq:E_kin_str_epsilon,eq:p_beta} and Taylor expansions at zero velocity one may write
\begin{equation}
E_{\text{kin}}(\beta) \approx \frac{1}{2}m_0c^2 \beta^{2} \left[ 1 + \frac{3}{4} \left(\frac{\beta}{\epsilon}\right)^{2}  + \frac{5}{8} \left(\frac{\beta}{\epsilon}\right)^{4}  + \mathcal{O}\left(\left(\frac{\beta}{\epsilon}\right)^{6}\right) \right] \, ,
\label{eq:E_kin_str_epsilon_taylor}
\end{equation}
\begin{equation} 
\vec{p} \,\approx\, m_0 c \vec{\beta}\,  \left[ 1 + \frac{1}{2}\left(\frac{\beta}{\epsilon}\right)^{2}  + \frac{3}{8} \left(\frac{\beta}{\epsilon}\right)^{4} + \mathcal{O}\left(\left(\frac{\beta}{\epsilon}\right)^{6}\right) \right] \, .
\label{eq:p_str_epsilon_taylor}
\end{equation}
These series are similar to \refp{eq:E_kin_exponential_taylor} and \refp{eq:p_exponential_taylor}; first two terms being the same. \Cref{eq:E_kin_str_epsilon,eq:p_beta} corresponding to \cref{eq:f_v_str_epsilon} are in agreement with the results of classical physics if $v \ll \epsilon c$  ($\beta \ll \epsilon$).

\section{\label{sec:comparison}Comparison of different alternatives}
\begin{figure*}[!htb]
\centering
\begin{subfigure}[t]{0.45\textwidth}
\includegraphics*[width=\textwidth]{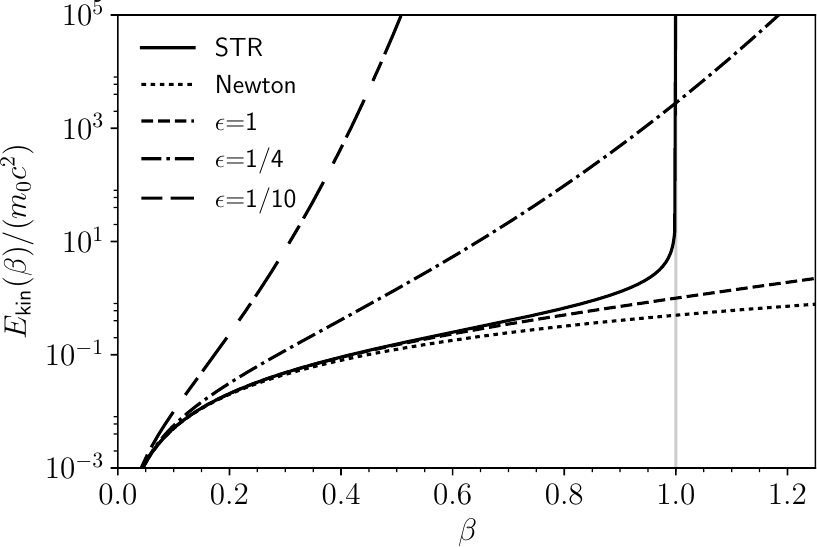}
\caption{\label{fig:exponential}Kinetic energy given by \cref{eq:E_kin_str_epsilon} for three values of parameter $\epsilon$ (corresponding to infinite maximal velocity $\beta_{\text{max}}$) compared to the dependence derived in STR and the dependence known in the classical physics of Newton (see \cref{eq:E_kin_newton}).
}
\end{subfigure}
\quad
\begin{subfigure}[t]{0.45\textwidth}
\includegraphics*[width=\textwidth]{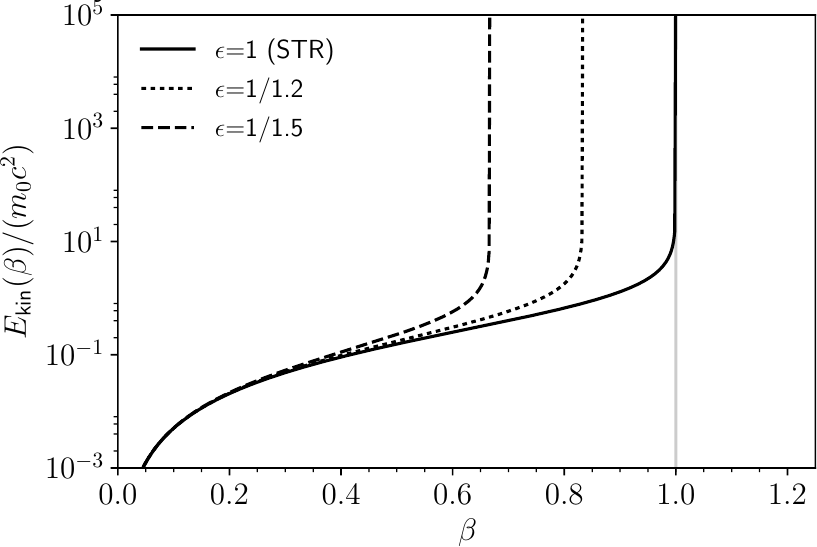}
\caption{\label{fig:STR}Kinetic energy given by \cref{eq:E_kin_exponential} for three values of parameter $\epsilon$ (corresponding to finite maximal velocity $\beta_{\text{max}}$); the case with $\epsilon\!=\!1$ ($\beta_{\text{max}}\!=\!1$) represents the dependence known from STR.
}
\end{subfigure}
\begin{minipage}[t]{1.\textwidth}
\caption{\label{fig:E_kin_comparison}Comparison of kinetic energy $\,E_{\text{kin}}(\beta)\,$, divided by $\beta$-independent constant $m_0c^2$, in dependence on velocity $\beta$ of moving particle in different alternatives.
}
\end{minipage}
\end{figure*}
In the preceding sections all results concerning inertia increase with rising velocity in the framework of generalized classical physics have been derived under the additional assumptions given by \cref{eq:f_conditions} and guaranteeing agreement with original results of classical physics at very low velocity values. The formulas showing the dependences of kinetic energy on particle velocity value at divers choices of function  $\;f(\beta)\,$, containing one free parameter $\epsilon$, have been presented. 

\Cref{fig:E_kin_comparison} shows then comparison of several possibilities of kinetic energy (divided by constant $m_0c^2$ for convenience) increase with increasing particle velocity $\beta$. 
In \cref{fig:exponential} the results obtained with the help of exponential $\,f(\beta)\,$ function (see \cref{eq:f_v_exponential}) are plotted for three different values of free parameter $\epsilon$; they are compared to the prediction derived in STR corresponding to function $f(\beta)$ given by \cref{eq:f_v_str_epsilon} for $\epsilon\!=\!1$, and also to the classical physics of Newton where $f(\beta)$ is independent of velocity (see \cref{eq:f_v_newton}). In the case of the exponential $\,f(\beta)\,$ function the maximal value of velocity and corresponding maximal value of kinetic energy has not been limited as one may see also from the plot.

The dependences of kinetic energy corresponding to the other type inertia increase specified by function $\,f(\beta)\,$ defined by \cref{eq:f_v_str_epsilon} are then shown in \cref{fig:STR} (for three different values of parameter $\epsilon$, too). In this case each of the alternatives corresponds to different finite maximal velocity value given by \cref{eq:v_max_str_epsilon}. The case corresponding to $\,\epsilon\!=\!1\,$ ($\beta_{\text{max}}\!=\!1$) represents the dependence known from the special theory of relativity, see 
\cref{sec:finite_velocity_limit}.

While at very low velocity values each of the plotted dependences of kinetic energy has been in agreement with classical physics of Newton (which may bee seen also from the plots) it may behave quite differently at high velocity values according to the value of free parameter (choice of function $f(\beta)$).

\section{\label{sec:conclusion}Conclusion}	 
In the preceding it has been shown that the phenomenon of inertial mass increase with rising velocity value may be derived also in the framework of classical physics based on standard Hamilton's equations \refp{eq:hamilton} if the force law proposed by Newton has been generalized; i.e., the time change of momentum having been given by corresponding force, see \cref{eq:F_dpdt,eq:p_v}. The values of particle inertia masses in corresponding Hamiltonian are to be represented by function $\,m(\beta)\,=\,m_0\, f\!(\beta)\,$ where $\,f(\beta)\,$ has been increasing dimensionless function of velocity $\beta$ ($\beta=v/c$, where $\,c\,$ is the speed of light in vacuum) .

Relations \refp{eq:E_kin_beta}, \refp{eq:p_beta} and \refp{eq:F_beta}, derived in this paper for the first time, represent general formulas of kinetic energy, momentum and force in dependence on velocity for a given and in principle \emph{arbitrary} function $\,f(\beta)\,$, i.e., inertia increase. The newly derived general formulas provide, therefore, basis for comparison of different possible alternatives to experimental data. Parameter $m_0$ may be determined in principle from the data of classical physics at very low velocities. The actual dependence of function $f(\beta)$ (which may contain some additional free parameters, too) should be, however, derived from data at higher velocity values.

The derived general formulas may describe very different physical results. Very different dependences of kinetic energy on velocity may be obtained at high velocity values as it has been shown for several cases under different assumptions in 
\cref{sec:alternatives}. 
Some examples have been presented for cases with infinite as well as finite maximal values of velocity in \cref{fig:E_kin_comparison}. It has been shown that the general formulas \refp{eq:E_kin_beta}, \refp{eq:p_beta} and \refp{eq:F_beta} may reproduce, in special cases, the results of classical physics of Newton as well as the predictions derived in the framework of the special theory of relativity, see 
\cref{sec:transition_newton,sec:finite_velocity_limit}. 
Numerical comparison of dependences of kinetic energy in dependence on velocity for different alternatives of function $\,f(\beta)\,$ have been shown for demonstration purposes in
\cref{sec:comparison}.

\section{Acknowledgment}
\textit{It is my (MVL) duty to thank Jaroslav Rau\v{s}, earlier member of Institute of plasma physics of the Czech Academy of Sciences, for decisive contribution  to presenting this paper. Several years ago he was very interested in my effort to analyze the problem of inertia increase in the framework of generalized classical physics. When I was fully unsuccessful in finding an analytical solution and regarded a numerical approach as quite insufficient  (and decided to put aside my effort in the given direction) he contributed fundamentally to further solution of the given problem (finding analytically solvable case with exponential function). The preceding results would not be available probably until now without this contribution of his.}


\footnotesize
\begin{thebibliography}{99}

\bibitem{Einstein1905_relativity}
A.~Einstein, "Zur Elektrodynamik bewegter Koerper", Annalen der Physik 17, Issue 10, 891-921 (1905).

\bibitem{Einstein1905_inertia}
A.~Einstein, "Ist die Traegheit eines Koerpers von seinem Energieinhalt abhaengig?", Annalen der Physik 18, Issue 13, 639-641, (1905).

\bibitem{Arnold1989}
V.~I.~Arnold, Mathematical Methods of Classical Mechanics, Springer-Verlag, ISBN 0-387-96890-3 (1989).

\end{thebibliography}
\end{document}